\documentclass[10pt,conference,]{IEEEtran} 
\IEEEoverridecommandlockouts

\usepackage[hidelinks]{hyperref}
\usepackage{cite}
\usepackage{amsmath,amssymb,amsfonts}
\usepackage{graphicx}
\usepackage{textcomp}
\usepackage{xcolor}
\def\BibTeX{{\rm B\kern-.05em{\sc i\kern-.025em b}\kern-.08em
    T\kern-.1667em\lower.7ex\hbox{E}\kern-.125emX}}
    
\usepackage{enumitem}
\usepackage{color}

\usepackage[most]{tcolorbox}
\usepackage{multicol}
\usepackage{makecell} 
\usepackage{array} 
\usepackage{booktabs}
\usepackage{url}
\usepackage{footmisc}
\usepackage{xspace}
\usepackage{algorithm}
\usepackage{algpseudocode}
\usepackage{multirow}
\usepackage{tcolorbox}
\usepackage{tablefootnote}
\usepackage[normalem]{ulem}

\newcommand{\tool}{\textit{STAIR}\xspace}

\newcommand{\peter}[1]{\textcolor{blue}{{\it [peter says: #1]}}}
\newcommand{\jinqiu}[1]{\textcolor{red}{{\it [Jinqiu says: #1]}}}

\newcommand{\ys}[1]{\textcolor{orange}{{\it [yisen says: #1]}}}

\newcommand{\phead}[1]{\vspace{1mm} \noindent {\bf #1}}
\usepackage{tikz}

\usepackage{tabularx}
\definecolor{mygreen}{HTML}{2DA44E}
\definecolor{findingsback}{RGB}{235,245,255}
\definecolor{findingsborder}{RGB}{102,178,255}
\newtcolorbox{findings}{
    enhanced,
    colback=findingsback,
    colframe=white,
    borderline west={3pt}{0pt}{findingsborder},
    boxrule=0pt,
    arc=0pt,
    sharp corners,
    left=2mm, right=2mm, top=1mm, bottom=1mm,
}

\begin{document}

\title{Reusing Past Repairs Through Hierarchical Trajectory Abstraction for Coding Agents}

\author{
\IEEEauthorblockN{Yisen Xu\textsuperscript{1},
Jiayuan Zhou\textsuperscript{2},
Ruiqi Pan\textsuperscript{3},
Tse-Hsun (Peter) Chen\textsuperscript{1}}
\IEEEauthorblockA{\textsuperscript{1}SPEAR Lab, Concordia University, Montreal, Canada\\
\textsuperscript{2}Waterloo Research Center, Huawei, Canada\\
\textsuperscript{3}Huawei Technologies Co., Ltd.\\
yisen.xu@mail.concordia.ca, 
jiayuan.zhou1@huawei.com, panruiqi@huawei.com, peterc@encs.concordia.ca
}

}


\maketitle

\begin{abstract}

Although LLM-driven repair agents can tackle complex, repository-level
issues, they treat every issue independently and discard the procedural
knowledge accumulated from previous repairs.
We introduce \tool, a framework that converts historical repair
trajectories into hierarchical, reusable plans that can be adapted to
steer future repairs.
Each past trajectory is transformed into a multi-level tree that ranges
from fine-grained diagnostic actions to high-level repair strategies,
encoding experience at several granularities.
When a new issue arrives, \tool selects relevant plan nodes from
multiple abstraction levels, tailors them into executable,
issue-specific plans, and supplies them to the agent through its prompt.
On SWE-bench Verified, \tool integrated with Lingxi reaches
81.2\% Pass@1 using MiniMax M2.5 and 79.2\% using GPT-5.
The generated plans also generalize across agents: without any code
change, they lift the Pass@1 of a structurally different agent,
mini-SWE-agent~v2, from 75.8\% to 81.0\%.
Ablation experiments further show that mixing multiple abstraction
levels surpasses any single level and that raw, unabstracted
trajectories transfer substantially worse.

\end{abstract}


\section{Introduction}


Software bugs are inevitable in modern software development, and resolving them consumes a substantial portion of developer effort~\cite{DBLP:journals/cacm/LientzST78,DBLP:conf/icse/GazzolaMM18}. In practice, experienced developers rarely approach each bug in isolation. Instead, they draw on accumulated debugging experience, reusing fault localization strategies, editing patterns, and verification procedures that have proven effective for similar problems in the past~\cite{DBLP:conf/icse/ZhongS15}. More importantly, this reuse is done at multiple levels simultaneously. A developer may recall both the specific API call that fixed a similar null pointer exception in the same repository, and the general strategy of tracing data flow backward from the failure site. Hence, effective problem-solving depends on integrating both fine-grained specifics and coarse-grained transferable patterns~\cite{DBLP:conf/iccbr/Branting97,DBLP:journals/aicom/AamodtP94}.

Recent advances in large language models (LLMs) have led to a new generation of autonomous repair agents that can resolve real-world repository-level issues~\cite{DBLP:conf/nips/YangJWLYNP24, DBLP:conf/issta/0002RFR24,livesweagent,Sonar}. These agents interact with codebases through tool calls, execute tests, and iteratively refine patches. However, most existing repair agents treat each issue as an independent task. After each run, they discard the procedural knowledge embedded in past repair trajectories, such as effective localization strategies and common failure modes. While some studies~\cite{DBLP:journals/corr/abs-2507-23361,lingxi,experepair} have begun to explore knowledge reuse for issue resolution, they face three key limitations.

First, existing approaches use an LLM to summarize historical repairs represented in two forms: issue--patch pairs~\cite{lingxi} and agent trajectories~\cite{DBLP:journals/corr/abs-2507-23361,experepair}. Summaries generated from issue--patch pairs cannot capture the intermediate reasoning that led to the fix, while summaries generated from trajectories condense the repair process without preserving how its reasoning, actions, and verification steps build on one another. 
Second, they represent each repair at a fixed level of detail, forcing a trade-off between concrete but non-transferable guidance and general but less actionable strategies~\cite{DBLP:conf/iccbr/Branting97,planninghi}.
Third, they insert the retrieved historical repair summaries into the prompt as-is, leaving the agent to bridge the gap between a past fix and the current repository on its own. Thus, these summaries are generic guidance rather than an issue-specific repair plan.


To address these limitations, we propose \tool, a framework that transforms successful historical repair trajectories into reusable procedural knowledge and adapts this knowledge to new issues. \tool collects these trajectories by running an existing repair agent on previously resolved issues, then processes them offline before any target repair begins. Rather than reducing each repair to a short textual summary, \tool segments each trajectory by repair stage and abstracts it into a multi-level hierarchy that captures how concrete agent actions connect to broader repair strategies. Lower levels of this hierarchy retain repository-specific details such as file paths and error messages, while higher levels capture transferable strategies that generalize across issues. Given a new issue, \tool retrieves relevant procedural knowledge across these levels and uses an LLM to adapt it to the target issue and repository, producing stage-specific plans. 

We evaluate \tool on SWE-bench Verified~\cite{jimenez2024swebench}, a benchmark of 500 real-world GitHub issues with developer-written test suites. \tool achieves a Pass@1 of 81.2\% with MiniMax M2.5 and 79.2\% with GPT-5. We also compare \tool with agents that reuse historical knowledge, including SWE-Exp~\cite{DBLP:journals/corr/abs-2507-23361}, Lingxi~\cite{lingxi}, and ExpeRepair~\cite{experepair}, and show that \tool consistently outperforms them. 

To assess generalizability, we transfer \tool's plans to another agent scaffold, mini-SWE-agent v2~\cite{DBLP:conf/nips/YangJWLYNP24} without any modification. The plans improve its Pass@1 from 75.8\% to 81.0\% while slightly reducing token usage, demonstrating that the learned plans capture reusable repair strategies rather than agent-specific heuristics. Our ablation study further shows that collapsing the hierarchy to any single abstraction level degrades performance substantially, with the raw-trajectory level alone yielding the largest drop. The findings further demonstrate the benefits of our multi-level abstraction.

In summary, our paper makes the following contributions:
\begin{itemize}
    \item We propose \tool, an approach that abstracts historical repair trajectories into a multi-level knowledge hierarchy. This hierarchy aggregates recurring repair patterns across similar issues at multiple granularities, enabling both repository-specific guidance and cross-issue transfer.
    \item \tool achieves 81.2\% Pass@1 on SWE-bench Verified. Ablation studies confirm that both the multi-level hierarchy and plan adaptation contribute to the observed gains.
    \item The generated plans transfer to a structurally different repair agent without any modification, improving its Pass@1 by 5.2\% while reducing token usage, confirming that the learned knowledge is agent-agnostic.
\end{itemize}

\phead{Paper Organization.}
Section \ref{sec:related} discusses related work. Section \ref{sec:methodology} details the design of \tool. Section \ref{sec:evaluation} presents the evaluation results. Section \ref{sec:threats} discusses potential threats to validity. Section \ref{sec:conclusion} concludes the work. 



\section{Related Work} \label{sec:related}

In this section, we review prior work most relevant to \tool in two
areas: LLM-based autonomous repair agents and trajectory-based learning and experience reuse in LLM agents.

\subsection{LLM-Based Autonomous Repair Agents}
\phead{Agent Architectures for Issue Resolution.} A growing body of work has explored how to equip LLMs with tool-use capabilities to resolve real-world repository-level issues. Early approaches demonstrate that even lightweight designs can be effective. 
Agentless~\cite{DBLP:journals/pacmse/XiaDDZ25} adopts a fixed and lightweight three-phase pipeline of fault localization, patch generation, and validation without any agent-driven planning. 
SWE-agent~\cite{DBLP:conf/nips/YangJWLYNP24} proposes an agent-computer interface that grants the model direct access to bash commands for navigating repositories, editing files, and running tests. AutoCodeRover~\cite{DBLP:conf/issta/0002RFR24} augments agents with AST-based code search to enable structured codebase exploration. 

As benchmarks become more challenging, recent agents adopt increasingly sophisticated strategies. OpenHands~\cite{DBLP:conf/iclr/0001LSXTZPSLSTL25} provides an open platform where agents write and execute code to resolve issues. Prometheus~\cite{pan2025prometheus} constructs a unified knowledge graph over the repository to support memory-enhanced navigation within a multi-agent system. TRAE~\cite{traeresearchteam2025traeagent} improves resolution rates through test-time scaling via modular solution generation, and live-SWE-agent~\cite{livesweagent} enables the agent to autonomously modify its own scaffold during runtime. Despite their strong results, these agents treat each issue as an independent task and discard all procedural knowledge after every run.

\phead{Early Attempts at Knowledge Reuse.}  
A few recent agents have begun to incorporate historical experience into the repair process. Lingxi~\cite{lingxi}, which serves as the base agent in our work, takes historical issue reports and their patches, uses an LLM to infer the reasoning behind each fix, and summarizes it into knowledge descriptions that are retrieved to guide issue resolution. SWE-Exp~\cite{DBLP:journals/corr/abs-2507-23361} uses an LLM to summarize each past repair trajectory independently into a natural-language description and retrieves relevant ones to augment the agent's prompt. ExpeRepair~\cite{experepair} stores successful repair cases from past trajectories and retrieves similar ones as reference examples when fixing a new issue. It appends general repair insights summarized by an LLM to the agent's prompt.

However, these approaches all inject retrieved knowledge directly into the agent's prompt without adapting it to the target issue, producing generic hints rather than issue-specific repair plans. They also represent knowledge at a single level of abstraction, unable to simultaneously provide concrete localization guidance and transferable high-level strategies. Moreover, none evaluates whether the resulting knowledge transfers across repair agents with different scaffolds. We address these limitations by constructing a multi-level knowledge hierarchy that aggregates recurring repair patterns across similar issues at multiple granularities and adapts them into stage-specific plans covering fault localization, planning, and code editing. We further show that the resulting plans transfer to a different repair agent without modification.

\subsection{Trajectory-Based Learning and Experience Reuse in LLM Agents}

\phead{Experience Extraction from Agent Trajectories.} Recent work has explored enabling LLM-based agents to 
improve from their own trajectories without parameter updates. Existing methods extract knowledge from trajectories in various forms. ExpeL~\cite{explaaai} distills trial-and-error experiences into natural-language insights. Reflexion~\cite{reflectionnips} generates verbal self-reflections from failed attempts to guide subsequent retries. AutoGuide~\cite{autoguidenips} derives conditional guidelines from offline trajectories to serve as context-aware prompts. These methods demonstrate that agent trajectories provide valuable guidance for future execution. 

However, the extracted knowledge is organized as a flat list, with all entries at the same level of abstraction and no association with repair stages. This leads to three limitations. First, the agent cannot distinguish between high-level strategies and low-level actions, so all guidance is treated as equally specific. Second, it cannot distinguish between guidance for fault localization and patch generation, making retrieval stage-agnostic and imprecise. Third, knowledge is derived from individual trajectories in isolation, without capturing patterns that generalize across issues. \tool addresses these limitations by organizing trajectory-derived knowledge into a hierarchical structure indexed by repair stage and aggregated across issues. This allows the agent to retrieve knowledge that matches the required abstraction level, repair stage, and recurring patterns observed across issues.


\phead{Workflow and Plan Induction from Trajectories.} More recent work moves toward inducing structured workflows or plans from agent trajectories. Agent Workflow Memory (AWM)~\cite{awmicml} identifies reusable workflows from successful web navigation trajectories and provides them as high-level guidance during future tasks. SAGE~\cite{sagecorr} applies a similar idea to software engineering. The agent abstracts a concise plan from its past trajectories and uses it to guide a refined second attempt on the same issue. SE-Agent~\cite{seagentcorr} revisits past trajectories and improves them through revision, recombination, and refinement. SWE-PRM~\cite{prmcorr} trains a process reward model that monitors trajectories during execution and intervenes when agents exhibit inefficient behaviors. 

These approaches show that structured trajectory reuse can improve both effectiveness and efficiency. However, prior studies focus on improving a single task. Namely, the agent learns from its own trajectory of the same issue and cannot be reused across issues. \tool builds on this idea and studies how procedural knowledge learned from prior issues transfers to new issues. 

\textbf{Positioning of Our Work.} Our work draws on the principles of case-based reasoning~\cite{DBLP:journals/aicom/AamodtP94}, particularly the idea that effective knowledge reuse requires hierarchical abstraction that spans multiple levels of granularity~\cite{DBLP:conf/ewcbr/BergmannW96,DBLP:conf/iccbr/Branting97}. Unlike prior trajectory-based learning methods that produce flat knowledge representations or single-level abstractions, \tool organizes historical repair trajectories into a multi-level abstraction tree ranging from concrete diagnostic steps to general repair strategies. Unlike SAGE and AWM, which improve performance on the same task through self-reflection, \tool learns reusable plans from previously resolved issues and transfers them to unseen issues. Furthermore, \tool adapts retrieved guidance into stage-specific plans tailored for the target issue. Our findings show that these plans transfer across structurally different repair agents and help achieve state-of-the-art results.


\begin{figure*}
  \centering
  \scalebox{1}{
  \includegraphics[width=1\textwidth]{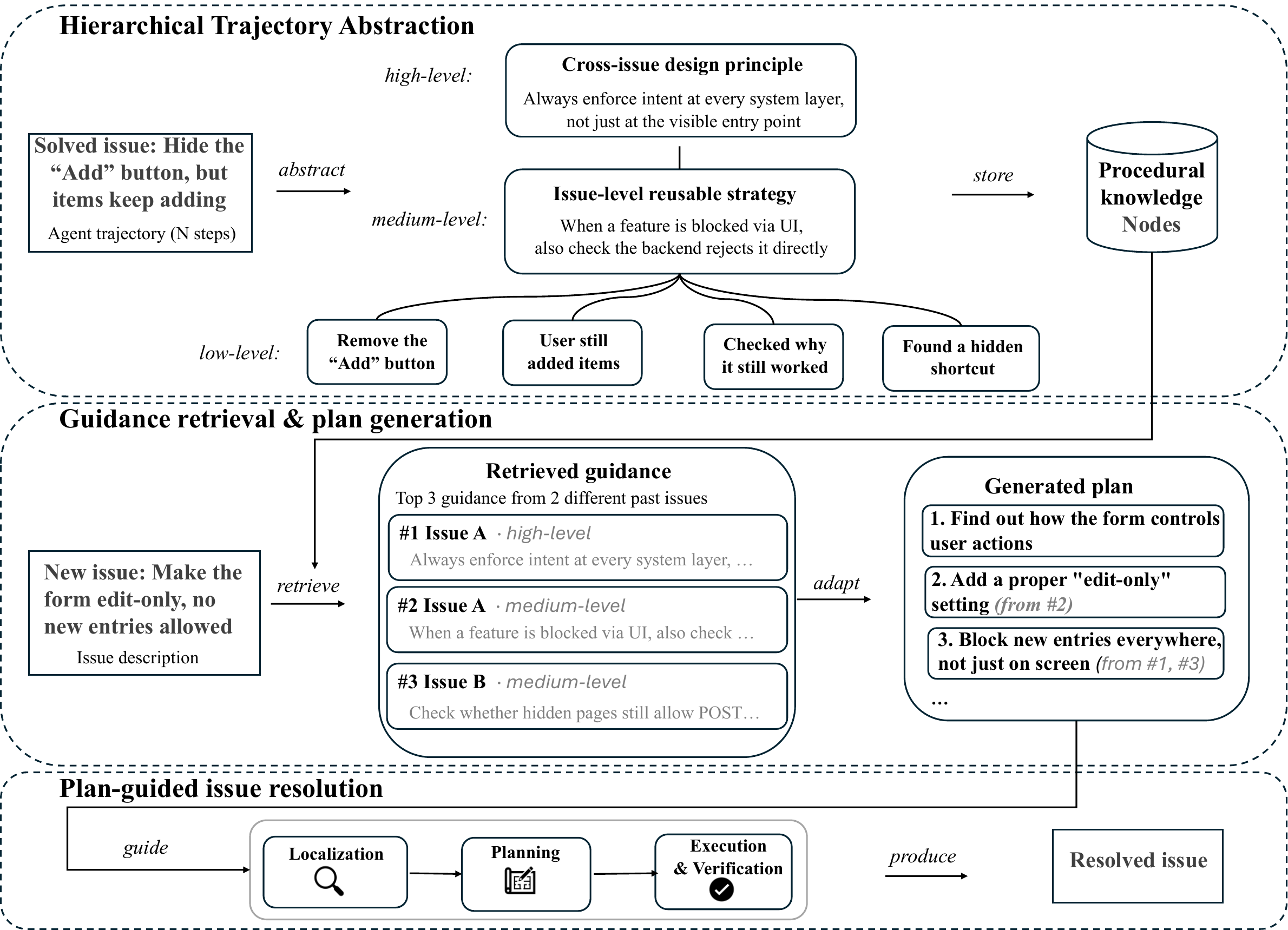}}
  \caption{An overview of \tool.}
  \label{fig:rag}
\end{figure*}


\section{Methodology}
\label{sec:methodology}

Figure~\ref{fig:rag} illustrates the overall workflow of \tool. During offline knowledge construction, \tool divides each successful historical repair trajectory into three stages: localization, planning, and execution-and-verification. It then abstracts the information within each stage at multiple levels, from concrete agent actions to broader procedural strategies. The resulting abstractions form a hierarchical repository of reusable procedural knowledge.

Given a new issue, \tool retrieves relevant procedural knowledge for each repair stage, potentially drawing different stages from different historical issues. The retrieved knowledge provides stage-specific \emph{guidance} that combines broadly transferable strategies with more actionable procedures. An LLM then adapts this guidance to the target issue and repository to generate issue-specific \emph{{plans}} for localization, planning, and execution-and-verification. These plans translate the retrieved procedural knowledge into concrete actions that the repair agent can follow.




\subsection{Procedural Knowledge Construction}
\label{subsec:memory}

STAIR collects historical issue-fixing trajectories by running Lingxi's repair scaffold without its knowledge-reuse mechanism on previously resolved issues. For each historical issue $u$, the system produces a trajectory $\tau(u)$ that captures the agent's reasoning, tool interactions, edit decisions, and verification feedback across the entire repair process.
To enable structured reuse, \tool organizes this trajectory in a stage-aware manner. Specifically, each trajectory $\tau(u)$ is partitioned into stage-specific segments $\tau_s(u)$ corresponding to localization, planning, and execution-and-verification~\cite{DBLP:journals/jss/LiuLKKLKB21,DBLP:journals/cacm/GouesPR19}. Each segment retains the same types of information (i.e. observation, thought, and action), but is restricted to the corresponding stage. This stage-wise decomposition bridges the rich, heterogeneous information in the full trajectory with the stage-specific abstraction and retrieval used later. 

In our implementation, to facilitate a controlled comparison with Lingxi~\cite{lingxi}, we follow its protocol for selecting historical issues. For each target issue, candidate historical issues are drawn from the same repository. The historical issues are restricted to those created before the target issue to avoid data leakage. They are then ranked by embedding similarity and filtered by an LLM-based relevance verifier. The repair trajectories of the selected historical issues are generated by running Lingxi without its knowledge-reuse mechanism. 

\subsection{Hierarchical Trajectory Abstraction}
\label{sec:abstraction}

Raw repair trajectories are often lengthy and noisy, containing fine-grained tool outputs, failed attempts, and redundant actions that are difficult to reuse directly. To improve reusability, \tool transforms each trajectory into a multi-level abstraction tree. 
This design reflects how developers work in practice: they move between high-level strategies and low-level actions, reusing past solutions at the level of detail that fits the current problem~\cite{DBLP:conf/ewcbr/BergmannW96,DBLP:conf/iccbr/Branting97,DBLP:journals/aicom/AamodtP94}. Accordingly, lower levels retain concrete, step-level actions, while higher levels capture more general repair strategies. We refer to the elements at each level of this tree as \emph{procedural knowledge nodes} (or simply \emph{nodes}), each capturing procedural knowledge at a given level of abstraction.

Within each stage of the repair workflow, \tool constructs a hierarchy of these procedural knowledge nodes. The leaf nodes correspond to individual trajectory steps, each containing the agent's thought, action, and observation. Each higher-level node summarizes a group of related nodes from the level below, representing broader procedures and repair strategies.



Formally, let $\mathcal{L}_0$ denote the leaf-node set constructed from the original trajectory steps, where each step contains the agent's thought, action, and observation. At abstraction level $\ell \ge 0$, an LLM-based grouping operator $G_\ell(\cdot)$ partitions $\mathcal{L}_\ell$ into $m_\ell$ non-overlapping groups, where each group $C_{\ell,j} \subseteq \mathcal{L}_\ell$:
\begin{equation} G_\ell(\mathcal{L}_\ell) = \{C_{\ell,1}, C_{\ell,2}, \ldots, C_{\ell,m_\ell}\}. 
\end{equation}
An LLM-based abstraction operator $A_\ell(\cdot)$ then abstracts each group into
a parent node, producing the next level:
\begin{equation}
\mathcal{L}_{\ell+1} =
\{A_\ell(C_{\ell,j}) \mid 1 \le j \le m_\ell\}.
\end{equation}
For instance, as shown in Figure~\ref{fig:rag}, the agent's trajectory for a solved issue includes several low-level steps: removing the ``Add'' button, discovering that users could still add items, investigating why, and finding a hidden shortcut. $G_\ell(\cdot)$ groups these steps because they share a common sub-goal of understanding why blocking the UI entry point did not prevent the undesired behavior. $A_\ell(\cdot)$ then abstracts the group into a medium-level strategy: \textit{when a feature is blocked via UI, also check that the backend rejects it directly}. At higher levels, this is further generalized into a cross-issue design principle: \textit{always enforce intent at every system layer, not just at the visible entry point}.

This process repeats until $|\mathcal{L}_{\ell+1}| \le K$, where $K$ is the number of root nodes (set to 2 in our experiments).
The resulting layers collectively form the hierarchical abstraction of the original trajectory. 
Since a repair trajectory naturally consists of distinct
phases (localization, planning, and execution with
verification), \tool first segments the full trajectory
$\tau(u)$ into stage-specific sub-trajectories
$\tau_s(u)$, and applies the above hierarchical
abstraction procedure independently to each one. This
preserves stage-level separation, enabling more targeted
retrieval during downstream guidance generation.

\begin{figure}[t]
\centering
\includegraphics[width=\columnwidth]{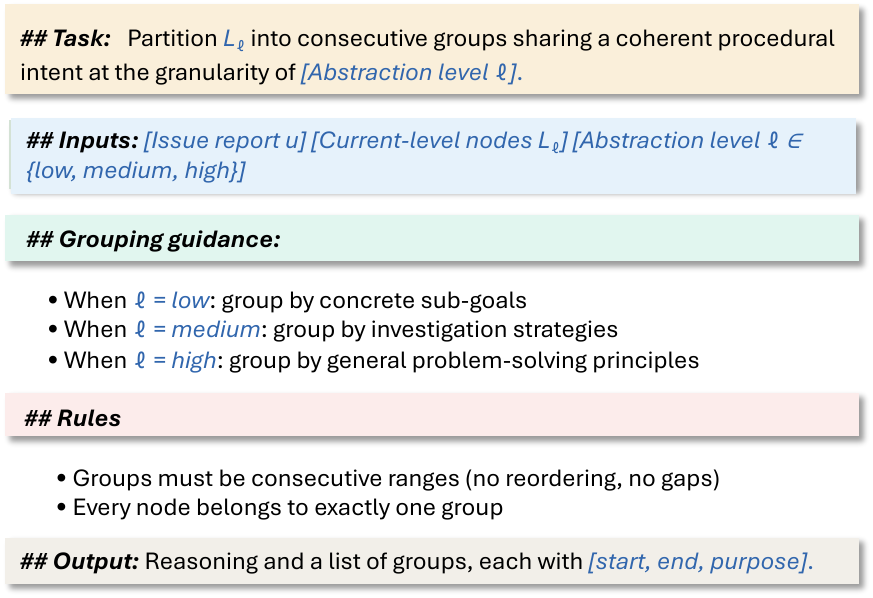}
\caption{Simplified prompt template for the grouping operator $G_\ell(\cdot)$. The same template is applied iteratively at each abstraction level. }
\vspace{-0.2mm}
\label{fig:prompt_group}
\end{figure}
\begin{figure}[t]
\centering
\includegraphics[width=\columnwidth]{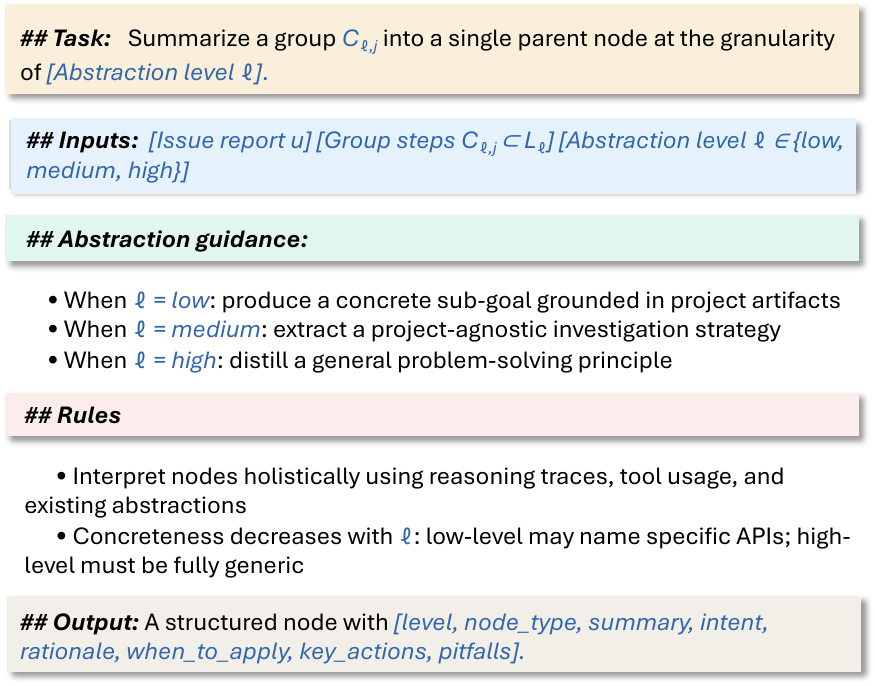}
\caption{Simplified prompt template for the abstraction operator $A_\ell(\cdot)$. The same template is applied iteratively at each abstraction level.}
\vspace{-0.2mm}
\label{fig:abstract_group}
\end{figure}

In our implementation, both the grouping operator $G_\ell(\cdot)$ and the
abstraction operator $A_\ell(\cdot)$ are implemented via LLM-based generation
using GPT-5~\cite{DBLP:journals/corr/abs-2601-03267}.
Starting from the raw trajectory steps $\mathcal{L}_0$, the two operators
are applied in alternation: $G_\ell(\cdot)$ partitions the current-level nodes
into consecutive groups that share a coherent procedural intent
(Figure~\ref{fig:prompt_group}), and $A_\ell(\cdot)$ abstracts each group into a
structured node containing the repair intent, key actions, applicable
conditions, and common pitfalls (Figure~\ref{fig:abstract_group}). The
output nodes $\mathcal{L}_{\ell+1}$ then become the input to the next
iteration. Before each iteration, an LLM-based decision prompt evaluates
the current nodes and selects the appropriate abstraction level~$\ell$
from three views: low-level for concrete sub-goals grounded in
project-specific artifacts, medium-level for project-agnostic
investigation strategies, and high-level for general problem-solving
principles that transfer across projects. While all levels share the same
node schema, the increasing abstraction naturally filters out
implementation details and surfaces reusable procedural knowledge, as
illustrated in Figure~\ref{fig:rag}.

\subsection{Guidance Retrieval and Plan Generation}
\label{sec:retrieve}
As described in Section~\ref{subsec:memory}, each repair trajectory is segmented into three stages: localization, planning, and execution with verification. To resolve a new issue, \tool retrieves relevant procedural knowledge from similar historical issues for each stage and organizes them into guidance, which is then adapted into issue-specific plans. We define these two concepts as follows. 

\smallskip\noindent\textbf{Guidance.}
For a target issue $u$ and repair stage $s$, the
\emph{guidance} $\mathcal{G}_s(u)$ is an ordered sequence of {\it abstract} procedural knowledge nodes retrieved from similar historical issues, arranged according to their original execution order in the source trajectories. Guidance captures historically effective repair procedures relevant to the target issue at stage $s$.

\smallskip\noindent\textbf{Plan.}
For a target issue $u$ and repair stage $s$, the
\emph{plan} $P_s(u)$ is a structured, issue-specific action sequence
generated by adapting the guidance $\mathcal{G}_s(u)$ to the concrete
symptoms and repository context of $u$.

\noindent\textit{\underline{Retrieval.}} Given a target issue $u$, \tool retrieves candidate procedural knowledge nodes from similar historical issues separately for each repair stage.
Each node $n$ is associated with a specific repair stage and can originate at any level of the abstraction hierarchy, as described in Section~\ref{sec:abstraction}, for similar historical issues.
This allows the retrieved guidance to combine low-level actionable details with high-level repair strategies.

Retrieval begins with a similarity-based scoring step that computes an initial relevance score $r(u,n)$ based on textual similarity matching between the issue report and node content. To refine this ranking, an LLM-based verifier evaluates the top candidates to assess their procedural relevance to the target issue, filtering out nodes that are lexically similar but not applicable in the current repair context. The retained nodes for stage $s$ are then ordered according to their execution sequence in the source trajectories, forming the guidance $\mathcal{G}_s(u)$, which is a coherent sequence that preserves the procedural structure of historical repair processes.

\noindent\textit{\underline{{Plan Generation.}}}
While the retrieved guidance captures useful repair procedures, it
remains abstract and does not yet specify concrete actions for the
current issue. \tool therefore adapts this guidance into executable
plans. For each stage $s$, \tool provides the LLM with the issue
report $u$ and the corresponding guidance $\mathcal{G}_s(u)$ and asks
it to generate a structured plan $P_s(u)$. The issue report conveys the
concrete symptoms and constraints of the current issue, while the
guidance supplies historically effective repair strategies. During plan
generation, the model selectively instantiates applicable procedures and
discards irrelevant ones, producing a plan that preserves useful procedures while grounding actions in the current repository. 

\subsection{Plan-Guided Issue Resolution}
\label{sec:guided_run}

The stage-specific plans generated in the previous step are executed to
resolve the target issue. For each repair stage $s$, the plan $P_s(u)$
provides structured procedural guidance that translates the retrieved repair knowledge into concrete actions while remaining adaptable to the
current repository context.

During localization, the plan $P_{\textit{loc}}(u)$ directs the agent toward the most relevant failure evidence, focuses analysis on suspicious regions of the codebase, and organizes diagnosis around a smaller set of plausible fault hypotheses, narrowing the search space from the full repository to a targeted set of candidate locations.
During planning, the plan $P_{\textit{plan}}(u)$ specifies the intended modification strategy, expected repair scope, and constraints that should guide subsequent edits. Based on this plan, the agent proposes candidate modifications and constructs a candidate patch. During execution-and-verification, the patch is validated through test-based checking, and the agent revises the patch when verification feedback indicates that the current repair attempt is insufficient.

Although the plans guide the repair process, they do not fully determine
it. The repair run still depends on evidence uncovered during
localization and feedback obtained during verification. Newly observed
evidence may refine the repair strategy, while failed verification may
require revisiting earlier decisions. In this way, historical procedural
knowledge informs the repair process without forcing it to follow a
rigid script.

\subsection{Using Existing Repair Agent as the Scaffold}
\label{sec:scaffold}

\tool is not tied to a particular repair agent and can augment agents that accept external repair guidance through their prompts. We use Lingxi~\cite{lingxi} as the underlying agent scaffold because it also reuses procedural knowledge from historical issues. This allows a controlled comparison between Lingxi's original knowledge-reuse mechanism and \tool's hierarchical procedural knowledge, while keeping the underlying repair workflow fixed. Lingxi also achieves competitive performance on SWE-bench Verified~\cite{jimenez2024swebench}, making it a strong underlying agent for our evaluation.

To integrate \tool, we retain Lingxi's internal decision logic, tools, and stage structure, disable its original knowledge-reuse mechanism, and insert the corresponding \tool-generated plans into the prompts for each stage at runtime.
To evaluate whether the generated plans transfer beyond Lingxi, we also integrate them with \textit{mini-SWE-agent~v2}~\cite{DBLP:conf/nips/YangJWLYNP24} in RQ2.

\section{Evaluation}
\label{sec:evaluation}

\subsection{{Benchmark and Metrics}} We conduct all experiments on \textbf{SWE-bench Verified}~\cite{jimenez2024swebench}, a human-curated subset of 500 instances from the original SWE-bench. Each instance corresponds to a real-world GitHub issue paired with a developer-written test suite that distinguishes correct patches. We use \textbf{Pass@1} as our evaluation metric, defined as the percentage of instances for which the LLM agent produces a correct patch on the first attempt. 

\subsection{{Models and Baselines}} We evaluate \tool using two state-of-the-art backbone models: \textbf{GPT-5}~\cite{DBLP:journals/corr/abs-2601-03267} (High Reasoning, commercial) and \textbf{MiniMax M2.5}~\cite{Minimax-m2.5} (High Reasoning, open-weight). For cross-scaffold transferability (RQ2), we apply the procedural plan generated by \tool to \textit{mini-SWE-agent v2}~\cite{DBLP:conf/nips/YangJWLYNP24}. 
We use the same settings and prompts across models and agent scaffolds to ensure a fair comparison.

\subsection{\textbf{Implementation Details}} 
We implement \tool using LangGraph~\cite{langgraph}, which orchestrates the entire hierarchical repair pipeline, including agent workflow coordination, trajectory storage, and stage-wise retrieval described in Section~\ref{sec:methodology}. The historical trajectories are collected from previously resolved issues by running Lingxi without its knowledge-reuse mechanism~\cite{lingxi}, and abstracted using the pipeline described in Section~\ref{subsec:memory}. To prevent temporal leakage, the historical issue candidate pool for each target instance is restricted to issues created before the target issue's creation time.

\subsection{{Environment}} As described in Section~\ref{sec:scaffold}, \tool uses Lingxi~\cite{lingxi} as the underlying repair agent. We follow Lingxi's agent configuration, including its tool definitions, prompting templates, and stage structure. All experiments are executed via API calls to GPT-5 and MiniMax M2.5. Resolving a single SWE-bench Verified instance takes 10–15 minutes on average, end-to-end. The per-instance cost of issue resolution, including retrieval, target-specific plan adaptation, analysis, patching, and verification, is \$0.84 with MiniMax M2.5 and \$1.60 with GPT-5. Hierarchical trajectory abstraction is constructed once using GPT-5 as an offline preprocessing step, and the resulting abstraction nodes are reused across all backbone LLMs. This preprocessing costs approximately \$0.44 per source trajectory.  In our experiments, \tool retrieves and abstracts 479 similar historical trajectories 
in total, resulting in a total preprocessing cost of approximately \$211. This cost is incurred once and amortized over all subsequent evaluations.

\subsection*{RQ1: How does \tool perform compared to other state-of-the-art autonomous repair agents?}
\label{sec:rq1}

\phead{Motivation and Approach.} 
The goal of this RQ is to evaluate whether \tool's hierarchical trajectory abstraction and plan adaptation improve repository-level issue-resolution rates compared to state-of-the-art repair agents. We compare \tool with nine existing agents on SWE-bench Verified, 
organized into three groups.

The first group uses the same backbone models as \tool 
(GPT-5 and MiniMax M2.5):
\begin{itemize}
    \item \textit{OpenHands}~\cite{DBLP:conf/iclr/0001LSXTZPSLSTL25}: 
   an open-source platform for AI agents that interact with the world by writing code, executing commands, and browsing the web.
    \item \textit{Prometheus-v1.2.1}~\cite{pan2025prometheus}: 
    a multi-agent platform that leverages unified 
    knowledge graphs to automatically bug fixing.
    \item \textit{mini-SWE-agent v2}~\cite{DBLP:conf/nips/YangJWLYNP24}: 
    a minimal bash-only agent scaffold developed by the SWE-bench team.
\end{itemize}

The second group includes other top-performing repair agents on the 
SWE-bench leaderboard that use different backbone models:
\begin{itemize}
    \item \textit{TRAE}~\cite{traeresearchteam2025traeagent}: 
   an ensemble reasoning agent that applies test-time scaling 
    through modular solution generation and selection.
    \item \textit{Sonar Foundation Agent}~\cite{Sonar}: 
    a single-agent tool-calling framework that combines bash execution with AST-based code search.
    \item \textit{live-SWE-agent}~\cite{livesweagent}: 
    a self-evolving agent that autonomously improves its own scaffold on-the-fly during runtime.
\end{itemize}

The third group includes experience reuse agents that retrieve 
knowledge from historical issues to guide repair:
\begin{itemize}
    \item \textit{SWE-Exp}~\cite{DBLP:journals/corr/abs-2507-23361}: 
    uses an LLM to summarize past repair trajectories into 
    natural-language descriptions and retrieves relevant ones 
    to augment the agent's prompt.
    \item \textit{Lingxi}~\cite{lingxi}: 
    takes historical issue reports and their patches, uses an LLM 
    to infer the reasoning behind each fix, and retrieves relevant 
    knowledge descriptions to guide issue resolution.
    \item \textit{ExpeRepair}~\cite{experepair}: 
    retrieves past repair examples from its own trajectories as 
    few-shot demonstrations and appends LLM-summarized repair 
    insights to the agent's prompt.
\end{itemize}




\begin{table}[t]
\centering
\caption{Comparison with state-of-the-art repair agents on SWE-bench Verified (Pass@1, \%). HR denotes High Reasoning.}
\label{tab:main_results}
\setlength{\tabcolsep}{3pt}
\scalebox{1.0}{
\begin{tabular}{@{}llr@{}}
\toprule
\textbf{Scaffold} & \textbf{Backbone Model} & \textbf{Pass@1 (\%)} \\
\midrule
\multicolumn{3}{@{}l}{\textit{Existing Agents (same backbone)}} \\
\textit{OpenHands}~\cite{DBLP:conf/iclr/0001LSXTZPSLSTL25} & GPT-5 & 71.8 \\
\textit{Prometheus-v1.2.1}~\cite{pan2025prometheus} & GPT-5 & 74.4 \\
\textit{mini-SWE-agent v2}~\cite{DBLP:conf/nips/YangJWLYNP24} & MiniMax M2.5 (HR) & 75.8 \\
\midrule
\multicolumn{3}{@{}l}{\textit{Existing Agents (other top repair agents)}} \\
\textit{TRAE}~\cite{traeresearchteam2025traeagent} & Doubao-Seed-Code & 78.8 \\
\textit{Sonar Foundation Agent}~\cite{Sonar} & Claude 4.5 Opus & 79.2 \\
\textit{live-SWE-agent}~\cite{livesweagent} & Claude 4.5 Opus (Medium) & 79.2 \\
\midrule
\multicolumn{3}{@{}l}{\textit{Experience Reuse Agents}} \\
\textit{SWE-Exp}~\cite{DBLP:journals/corr/abs-2507-23361} & Claude 4 Sonnet & 73.0 \\
\textit{ExpeRepair}~\cite{experepair} & Claude 4 Sonnet + o4-mini & 74.6 \\
\textit{Lingxi}~\cite{lingxi} & Claude 4 Sonnet & 74.6 \\
\textit{Lingxi} & MiniMax M2.5 (HR) & 74.6 \\

\textit{Lingxi} & GPT-5 (HR) & 75.6 \\

\midrule
\multicolumn{3}{@{}l}{\textit{Our Agents}} \\
\textbf{\textsc{Stair$_{GPT5}$}} & \textbf{GPT-5 (HR)} & \textbf{79.2} \\
\textbf{\textsc{Stair$_{MiniMax}$}} & \textbf{MiniMax M2.5 (HR)} & \textbf{81.2} \\
\bottomrule
\end{tabular}
}
\vspace{-2mm}
\end{table}

\phead{Results.} Table~\ref{tab:main_results} presents the performance of \tool alongside state-of-the-art agents on SWE-bench Verified. 
\textsc{Stair$_{MiniMax}$} achieves a Pass@1 of \textbf{81.2\%}, outperforming all other repair agents. In comparison, \textsc{Stair$_{GPT5}$} achieves a Pass@1 of 79.2\%, ranking second only behind \textsc{Stair$_{MiniMax}$}. This finding shows that \tool consistently delivers strong performance across different backbone models. 

When compared with repair agents that use the same backbone models, \tool achieves higher resolved rate. On GPT-5, \tool (79.2\%) outperforms \textit{OpenHands} (71.8\%) by 7.4\% and \textit{Prometheus-v1.2.1} (74.4\%) by 4.8\%. On MiniMax M2.5, \tool (81.2\%) exceeds \textit{mini-SWE-agent v2} (75.8\%) by 5.4\%.

Among other top repair agents, \textit{live-SWE-agent} and \textit{Sonar Foundation Agent} achieve the highest Pass@1 of 79.2\%, both built on Claude~4.5 Opus, one of the strongest backbones used across all prior repair agents~\cite{livesweagent,DBLP:conf/nips/YangJWLYNP24}. \textit{TRAE} (Doubao-Seed-Code) reaches 78.8\%. \tool, by contrast, surpasses all baselines while relying on comparatively weaker backbones: MiniMax and GPT-5. We chose these backbones to assess \tool's contribution independently of model capacity. Nonetheless, \tool outperforms the strongest Claude-based baselines, indicating that the observed gains are from the proposed framework rather than from the backbone choice.  

Compared with experience-reuse agents that also leverage historical issues, \tool outperforms all three across every backbone configuration. On Claude 4 Sonnet, SWE-Exp achieves 73.0\%, and both ExpeRepair and Lingxi achieve 74.6\%. To enable a controlled comparison on the same backbone LLM, we also run Lingxi (with its own procedural knowledge) with MiniMax M2.5 and GPT-5, obtaining 74.6\% and 75.6\%, respectively. \tool outperforms these same-backbone Lingxi baselines by 6.6\% (81.2\% vs.\ 74.6\% on MiniMax M2.5) and 3.6\% (79.2\% vs.\ 75.6\% on GPT-5), confirming that \tool's procedural knowledge provides more effective guidance than Lingxi's.

\begin{findings}

\tool achieves a Pass@1 of 81.2\% with MiniMax M2.5 and 79.2\% with GPT-5 on SWE-bench Verified, achieving top performance among compared agents under reported settings.

\end{findings}

\subsection*{RQ2: Can \tool's plans transfer to other repair agents?}
\label{sec:rq2}

\phead{Motivation and Approach.}
\tool learns and constructs procedural knowledge within its own agent scaffold, raising a key question: does it capture generalizable repair strategies, or does it encode scaffold-specific heuristics? 
To answer this, we transfer the stage-specific plans distilled from \tool's historical trajectories into \textit{mini-SWE-agent v2}~\cite{DBLP:conf/nips/YangJWLYNP24}, a repair agent with a different architecture, tool interface, and prompting strategy. 
\textit{mini-SWE-agent v2} is a widely adopted and lightweight repair agent developed by the SWE-bench team. It achieves over 74\% on SWE-bench Verified with a minimal architecture that relies solely on bash, making it an ideal candidate for evaluating whether procedural plans generalize across structurally different repair agents.

We prepend an \texttt{<external\_reference\_plans>} block to the agent's input at each repair stage, while leaving \textit{mini-SWE-agent v2}'s source code, tool definitions, and output verification logic unchanged. We use MiniMax M2.5 as the backbone model, as it achieves the best performance in RQ1 (Table~\ref{tab:main_results}).


\begin{table}
\centering
\small
\caption{Transferability of \tool's procedural plans to
  \textit{mini-SWE-agent v2} on SWE-bench Verified (500~instances).
  \textit{Avg. Steps} and \textit{Avg. token} denotes the average number of steps the agent took and the total token usage per instance.}
\label{tab:transfer}
\setlength{\tabcolsep}{6pt} 
\begin{tabular}{lrccc}

\toprule
\textbf{Configuration}
  & \textbf{Resolved}
  & \textbf{Avg. steps}
  & \textbf{Avg. token}\\
\midrule
\textit{mini-SWE-agent v2}
  & 379/500 (75.8\%)
  & \multicolumn{1}{c}{60.45}
  &1,244,923 \\
~~+ \tool Plan
  & 405/500 (81.0\%)
  & 58.18
  & 1,179,885 \\
\bottomrule
\end{tabular}
\vspace{-2mm}
\end{table}

\phead{Results.} 
Table~\ref{tab:transfer} reports the transfer results. Injecting \tool's procedural plans improves the Pass@1 of \textit{mini-SWE-agent v2} from 75.8\% (379/500) to 81.0\% (405/500), an absolute gain of 5.2\%. 
At the same time, the average token usage \textit{decreases} from 1,245k to 1,180k per instance. This reduction is driven by fewer agent steps (average steps decreased from 60.45 to 58.18), as the injected plans guide the agent toward more targeted exploration and reduce unnecessary search. 


These results demonstrate that procedural plans learned from \tool transfer effectively across repair agents without any modification (the plans are only prepended to the prompt). Despite being derived from a different agent, the plans improve both effectiveness (higher Pass@1) and efficiency (lower token usage), suggesting that they capture reusable repair strategies rather than agent-specific heuristics.

\begin{findings}
\tool's procedural plans are agent-agnostic, improving \textit{mini-SWE-agent v2}'s Pass@1 by 5.2\% (75.8\% to 81.0\%) while slightly reducing average token usage per instance due to fewer agent steps. 
\end{findings}

\subsection*{RQ3: How does plan guidance affect agent behavior?}

\phead{Motivation and Approach.}
Procedural plans can improve repair outcomes, but their role in agent behavior remains unclear. In particular, we seek to understand how plans contribute to successful repairs and, in regressed cases (i.e., previously passed issue becomes failed),
whether the failures are attributable to the plan itself or to downstream execution errors.  To study these questions, we manually analyze gained and regressed instances for two comparisons: \tool vs.\ \textit{Lingxi} (which uses its own knowledge-reuse mechanism), and \textit{mini-SWE-agent v2} with \tool's plans vs.\ vanilla \textit{mini-SWE-agent v2}.
For each instance, two authors independently examined the generated plan, the produced patch, and the full agent trajectory, then discussed disagreements to reach a consensus label. The resulting inter-rater agreement achieved a Cohen's kappa of 0.85, indicating a strong agreement~\cite{cohen1960coefficient}. Finally, we compare each plan-augmented run with its corresponding baseline to assess how the plan affected the agent's behavior.


\begin{table}[t]
\centering
\caption{Gain/regression distribution on SWE-bench Verified after adding \tool's generated plans. 
\emph{Shared} = solved by both; \emph{Gained} = solved only with \tool's
plan; \emph{Regressed} = solved only by baseline. Baselines are \textit{Lingxi} for \tool and vanilla \textit{mini-SWE-agent v2} for \textit{mini-SWE-agent v2} + Plans.}
\label{tab:gain-regress}
\setlength{\tabcolsep}{4pt}
\scalebox{0.95}{
\begin{tabular}{@{}l rrrr@{}}
\toprule
\textbf{Comparison} & \textbf{Shared} & \textbf{Gained} & \textbf{Regressed} & \textbf{Net} \\
\midrule
\textsc{Stair$_{MiniMax}$} vs.\ \textsc{Lingxi$_{MiniMax}$} & 363 & 43 & 10 & +33 \\
\textit{mini-SWE-agent v2} + Plans vs.\ vanilla & 372 & 33 & 7 & +26 \\
\bottomrule
\end{tabular}
}
\end{table}

\phead{Results.}
Table~\ref{tab:gain-regress} summarizes the gain/regression breakdown. In both repair agents, gains after adding \tool's plans substantially outnumber regressions (43 vs.\ 10 and 33 vs.\ 7), confirming that the plans provide a strong net benefit. We next analyze the two groups separately to understand the mechanisms behind these outcomes. 

\phead{Why Gained.}
We group the instances into three categories according to the plan's main contribution. Each category reflects a different problem that the plan corrects in the baseline behavior.

\begin{itemize}[leftmargin=1.5em]
\item \textbf{Original Fix Targets the Wrong Fault Location (Wrong Location)}: the baseline targets the wrong 
component. Without \tool's plan, the agent cannot recover from this early mistake and all subsequent steps are wasted. The plan guides the agent to the correct location from the start (e.g., fixing the upstream serializer rather than the downstream loader).

\item \textbf{Original Fix Covers Only Part of the Affected Code (Partial Fix)}: the baseline identifies the correct root cause but modifies only a subset of the relevant code locations. Adding the plan ensures all affected components are covered. For 
example, in \texttt{django-11138}, a timezone conversion bug affected 
MySQL, Oracle, and SQLite backends, but the baseline only fixed the 
first two. The plan framed it as a cross-backend audit, pushing the 
agent to cover all three.

\item \textbf{Original Fix Modifies More Code Than Necessary (Over-modification)}: the baseline applies a fix that changes more code than required, potentially introducing regressions. Adding the plan restricts the repair to the minimal necessary change (e.g., extending a matching condition instead of rewriting the entire discovery pipeline). 
\end{itemize}
 \begin{figure}[t]
\centering
\includegraphics[width=\columnwidth]{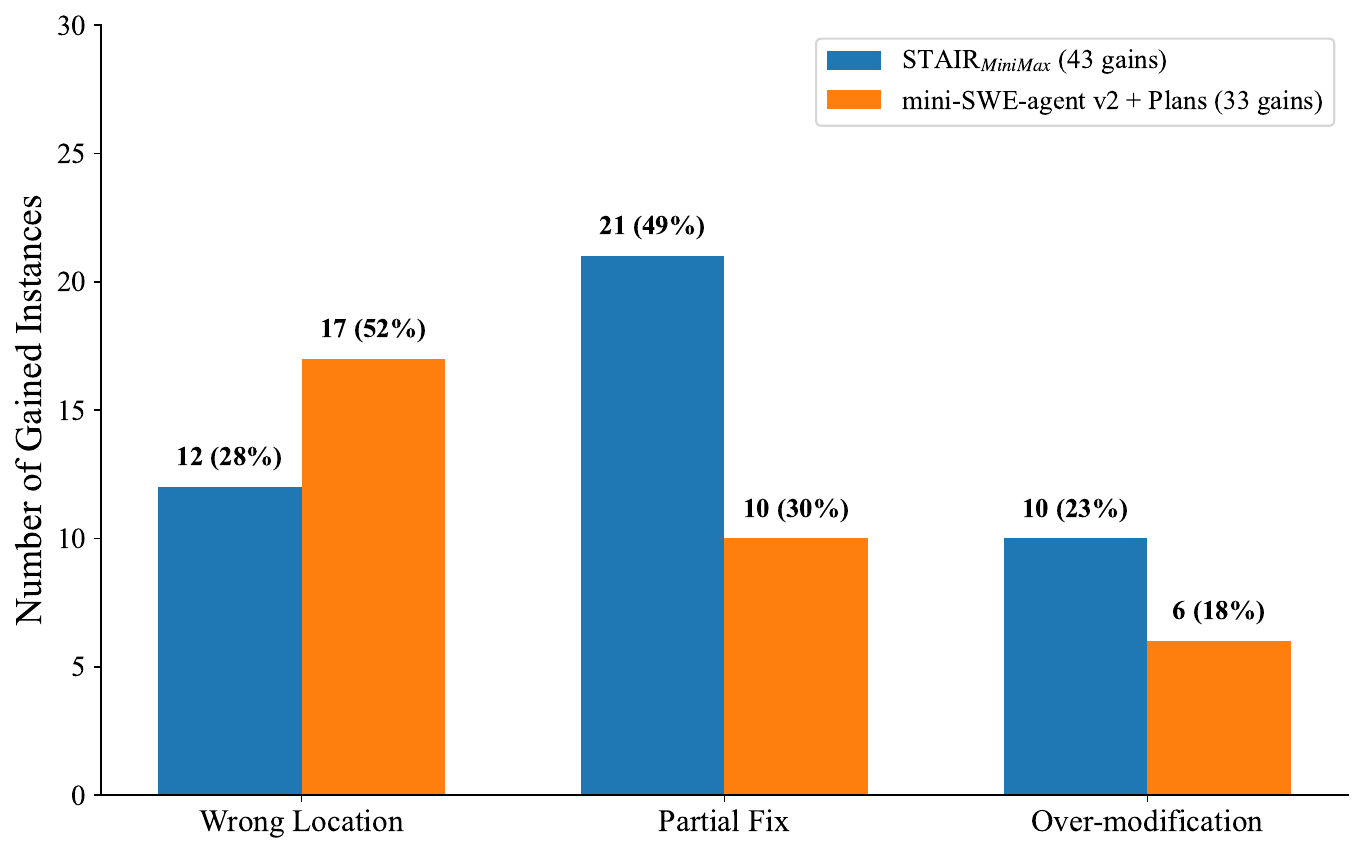}
\vspace{-0.5cm}
\caption{Distribution of gain categories for \textsc{Stair$_{MiniMax}$} and \textit{mini-SWE-agent v2} + Plans. }
\vspace{-0.2mm}
\label{fig:gain-breakdown}
\end{figure}

Figure~\ref{fig:gain-breakdown} shows the distribution of gain categories.
For \textsc{Stair$_{MiniMax}$} (43 instances), \textit{Partial Fix} dominates (49\%), followed by \textit{Wrong Location} (28\%) and \textit{Over-modification} (23\%).
For \textit{mini-SWE-agent v2} + Plans (33 instances), \textit{Wrong Location} dominates (52\%), followed by \textit{Partial Fix} (30\%) and \textit{Over-modification} (18\%).

This difference is also associated with the different workflows of the two agents. For \tool, the largest share of gains
comes from \textit{Partial Fix}, suggesting that the agents in \tool often reach the correct repair direction but benefit from plans that help it cover all affected locations. For \textit{mini-SWE-agent v2}, the largest share of gains comes from \textit{Wrong Location}, suggesting that plans are most useful in helping the agent focus on the correct problem early.

\phead{Why Regressed.}
We manually analyze all 17 regressed instances (10 from \textsc{Stair$_{MiniMax}$} and 7 from \textit{mini-SWE-agent v2} + Plans). We find that regressions fall into two categories: \emph{plan-related} failures, where the injected plan provides incomplete coverage of the required changes, and \emph{non-plan-related} failures, where the plan is reasonable but the agent fails during execution.

In the first case (8/17), the procedural plan captures the main repair direction
but does not fully specify the complete scope of changes needed
for a correct fix. This typically happens when the issue
requires coordinated updates across multiple files, entry
points, while the plan identifies only the
primary fix location. Across both agents, 8 out of 17 regressions fall into this category (5 from \textsc{Stair$_{MiniMax}$} and 3 from
\textit{mini-SWE-agent v2} + Plans).

In the second case (9/17), the procedural plan matches the issue, but the agent fails to follow it correctly during
execution. These failures include deviating from the intended
plan, introducing unnecessarily complex edits, or making local
implementation mistakes such as using the wrong API or passing
incorrect arguments. The remaining 9 regressions fall into this category (5 from
\textsc{Stair$_{MiniMax}$} and 4 from \textit{mini-SWE-agent v2}
+ Plans).

These findings suggest that the procedural plans are largely effective, with only a few that miss part of the required fix scope.  

\begin{findings}
Plans improve repair mainly by correcting fault localization,
expanding repair coverage, and reducing unnecessary edits.
Regressions, in contrast, arise either from incomplete plan
coverage or from execution mistakes that are not caused by the
plan itself.
\end{findings}

\subsection*{RQ4: What is the individual contribution of each component in \tool?}
\label{sec:rq4}

\phead{Motivation and Approach.} 
\tool's procedural knowledge is constructed through hierarchical trajectory abstraction, where raw repair trajectories are grouped into procedural segments and abstracted into nodes at multiple levels. This hierarchy captures both concrete step-level actions and high-level repair strategies. 

To evaluate the contribution of this abstraction, we conduct an ablation study along two dimensions: (1) raw trajectories vs.\ abstracted trajectories, and (2) different abstraction levels (low-level, medium-level, high-level, and multi-level). 
We implement the following variants:

\begin{itemize}
\item \textbf{No Abstraction (Raw Trajectories):} directly uses unprocessed trajectory logs without hierarchical abstraction.

\item \textbf{Low-Level Abstraction Only:} uses only fine-grained nodes that preserve concrete, repository-specific details (e.g., file paths, function signatures, error messages).

\item \textbf{Medium-Level Abstraction Only:} uses only coarse-grained nodes that capture general repair strategies independent of specific repositories.

\item \textbf{High-Level Abstraction Only:} uses only the most abstract nodes that capture general problem-solving principles transferable across projects.

\item \textbf{Full Hierarchical Abstraction (Full System):} combines nodes from multiple abstraction levels, reflecting the full \tool.
\end{itemize}

Due to the cost of running full-benchmark evaluations for each ablation variant, we conduct the study on a difficulty-stratified subset of 125 instances rather than all 500. We allocate approximately three-quarters of the subset (92 instances, 73.6\%) to the medium-and-hard tiers ($\geq$15-minute estimated fix time as annotated by SWE-bench Verified) and one-quarter (33 instances, 26.4\%) to the easy tier, following a roughly 3:1 ratio between non-trivial and trivial instances. This sampling strategy concentrates analytical power on instances where design choices are most likely to produce observable behavioral differences.  
All experiments use MiniMax M2.5 (High Reasoning) as the backbone.

\begin{table}
\centering
\small
\caption{Ablation study on a difficulty-stratified subset of 125 SWE-bench Verified instances. $\Delta$ denotes the absolute difference from the full hierarchical abstraction.}
\label{tab:ablation}
\setlength{\tabcolsep}{4pt}
\renewcommand{\arraystretch}{1.1}
\begin{tabular}{lcc}
\toprule
\textbf{Variant} & \textbf{Resolved (\%)} & $\Delta$ \\
\midrule
\textbf{Full Hierarchical Abstraction} & \textbf{100/125 (80.0)} & -- \\
\midrule
\multicolumn{3}{l}{\textit{Baseline (No Abstraction)}} \\
Raw Trajectories & 72/125 (57.6) & $-$22.4 \\
\midrule
\multicolumn{3}{l}{\textit{Ablations (Abstraction Level)}} \\
Low-Level only & 80/125 (64.0) & $-$16.0 \\
Medium-Level only & 76/125 (60.8) & $-$19.2 \\
High-Level only & 73/125 (58.4) & $-$21.6 \\
\bottomrule
\end{tabular}
\end{table}

\phead{Results.} Table~\ref{tab:ablation} summarizes the ablation results along two dimensions. Replacing hierarchical abstraction with raw trajectories reduces the resolution rate from 80.0\% to 57.6\% (-22.4\%). The findings indicate that raw trajectories (which may contain redundant actions, failed exploration paths, and issue-specific artifacts) do not transfer across issues. \tool's abstraction filters out such noise and restructures trajectories into reusable procedural knowledge, substantially improving generalization.

\phead{Different abstraction levels capture complementary aspects of procedural knowledge.} Using a single abstraction level consistently underperforms the full hierarchical design. Low-level-only achieves 64.0\% ($\Delta$=$-$16.0\%), medium-level-only 60.8\% ($\Delta$=$-$19.2\%), and high-level-only 58.4\% ($\Delta$=$-$21.6\%).

This trend reveals a clear progression: low-level abstractions are the most effective among single-level variants because they retain concrete, executable details needed for repository-specific edits. High-level abstractions perform the worst, as they capture only coarse strategies and lack grounding for direct action. Medium-level abstractions fall in between. Our findings show that effective repair requires procedural knowledge at multiple levels of abstraction, and collapsing to any single level leads to incomplete guidance and degraded performance.

\begin{findings}
Using raw trajectories without abstraction yields the largest performance drop ($-$22.4\% in Pass@1), showing that issue-specific actions in the raw trajectories do not generalize well to new issues. 
The full hierarchical abstraction outperforms all single-level variants, demonstrating that integrating multiple levels of procedural knowledge is necessary for effective repair.
\end{findings}



\section{Threats to Validity}
\label{sec:threats}


\subsection{Internal Validity} To reduce the potential biases in manual analysis, two authors independently labeled all gained and regressed instances and resolved disagreements by consensus, achieving a Cohen's kappa of 0.85, indicating strong agreement. 
The hierarchical abstraction process relies on LLM-based generation, including both the grouping and abstraction operators. As a result, the quality of the resulting multi-level nodes may vary with the complexity and length of the input trajectory. The plan compatibility filter also uses an LLM-based evaluator and may occasionally discard useful plans or retain inapplicable ones. 
The transfer results in RQ2 show that the extracted plans improve a different repair agent, indicating that the approach does not overfit to a specific agent or workflow.

\subsection{External Validity} Our evaluation is conducted on SWE-bench Verified (500 Python instances from 12 repositories), so empirical results may not directly generalize to other languages, benchmarks, or proprietary codebases. However, the framework is not inherently tied to SWE-bench Verified: it applies to any issue-resolution setting that provides executable tests for patch validation. The procedural knowledge is constructed from trajectories collected by Lingxi, and a different base agent may produce trajectories of different quality, potentially affecting the resulting plans. That said, the successful transfer of plans to mini-SWE-agent v2 in RQ2, without any modification, suggests that the plans capture agent-agnostic repair strategies rather than scaffold-specific artifacts.

\subsection{Construct Validity} Our study evaluates the effectiveness of \tool using Pass@1 and token consumption as primary metrics. While Pass@1 reflects whether a generated patch passes the test suite, it does not guarantee full semantic correctness beyond the provided tests. This limitation may allow behaviorally incorrect patches to be accepted if the tests do not fully capture the intended behavior~\cite{strengthicse,cureworse,overfittingse}. The underlying LLMs may affect the results. However, we conducted our experiments on one commercial and one open weight model (GPT-5 and MiniMax M2.5) and we found that the results are consistent.
\section{Conclusion}
\label{sec:conclusion}

In this paper, we introduced \tool, a framework that abstracts historical repair trajectories into multi-level hierarchical representations and adapts them into reusable plans to guide future repairs. \tool abstracts raw repair trajectories into multi-level representations ranging from concrete diagnostic steps to general repair strategies, retrieves relevant guidance for each stage of the repair workflow, and adapts it into issue-specific executable plans.

We evaluated \tool on SWE-bench Verified, where it achieves a Pass@1 of 81.2\% with MiniMax M2.5 and 79.2\% with GPT-5. An ablation study shows that combining multiple abstraction levels outperforms any single level, and that raw trajectories without abstraction generalize poorly, with the largest performance drop of 22.4\%. The generated plans also transfer to mini-SWE-agent v2 without any modification, improving its Pass@1 from 75.8\% to 81.0\% while reducing token usage, demonstrating that the learned knowledge is agent-agnostic.
More broadly, our results show that the effectiveness of experience reuse depends less on the source of historical knowledge than on how it is structured and applied. Raw trajectories and flat summaries both contain useful information, yet they transfer poorly when injected as-is. Organizing knowledge into multiple abstraction levels and adapting it into stage-specific plans bridges this gap, turning historical experience into actionable guidance that generalizes across issues, models, and agent scaffolds.

\bibliographystyle{IEEEtran}
\bibliography{sample-base}

\end{document}